# Strategies for Collecting Multi-Institutional Data in Discipline-Based Education Research

**Authors**: Meagan Sundstrom (Postdoctoral Researcher[1]), David Esparza (Postdoctoral Fellow[1]), Justin Gambrell (Fixed Term Assistant Professor[2]), Adrienne Traxler (Associate Professor[3]), and Eric Brewe (Professor[1*])

**Institutional Affiliations**:
[1]Department of Physics, Drexel University, Philadelphia, Pennsylvania 19104, USA
[2]Department of Computational Mathematics, Science and Engineering, Michigan State University, East Lansing, Michigan 48824, USA
[3]Department of Science Education, University of Copenhagen, Copenhagen, Denmark

***Corresponding Author**
Eric Brewe
Disque Hall 916, 3141 Chestnut St., Philadelphia, PA 19104, USA
305-457-7484
eb573@drexel.edu

**Manuscript Type**: Research Methods

**Characters**: 54,454 (With spaces; Excluding Abstract, Acknowledgements, and References)

**Running Title**: Multi-Institutional Data Collection

**Keywords**: Multi-institutional research; Institutional Review Board; Participant recruitment; Data collection; Research methods; Undergraduate

**Abstract**

Multi-institutional studies are critical for advancing discipline-based education research (DBER) because they allow us to determine where and for whom research findings are applicable. Despite this benefit, such studies remain relatively rare due to the complexities of coordinating data collection across different institutions. In this paper, we describe key challenges and propose actionable strategies for implementing multi-institutional DBER studies. We focus on navigating Institutional Review Board procedures, recruiting participants from a range of institution types, standardizing data sources across institutions, and managing logistics. We also provide an applied example of these strategies from a national research project in which we collected concept inventory data, social network surveys, and classroom observations from 31 introductory physics instructors at 28 institutions in the United States.

## INTRODUCTION

Several large-scale discipline-based education research (DBER) studies over the past two decades have examined instructional practices and student outcomes across institutions. These efforts range from landmark studies characterizing science teaching practices at scale (e.g., Stains *et al.*, 2018) to discipline-specific multi-institutional work in chemistry education (e.g., Barbera *et al.*, 2013; Naibert *et al.*, 2021; Van Dusen *et al.*, 2021), physics education (e.g., Sundstrom *et al.*, 2025a; Hazari *et al.*, 2010; Morley *et al.*, 2023; Zwickl et al., 2014), biology education (e.g., Andrews *et al.*, 2011; Couch *et al.*, 2019; Courtney *et al*., 2025; Ferrare *et al*., 2019), and engineering education (e.g., Huerta-Manzanilla *et al.*, 2021; Lord *et al.*, 2014; Yasuhara *et al.*, 2012; Wang *et al.*, 2022) research. Multi-institutional studies are necessary in DBER to strengthen the validity of findings; that is, to ensure that results are not solely applicable to a single instructional context (Tipton and Olsen, 2018). For studies that measure the impacts of educational interventions on student outcomes, multi-institutional data allow us to determine where and for whom these interventions are most effective and, in some cases, identify generalizable interventions that improve student outcomes across diverse instructional contexts.

Despite these affordances, multi-institutional studies remain relatively uncommon in the current DBER literature (National Research Council, 2012; Docktor and Mestre, 2014). Multi-institutional studies that contain near-complete datasets (e.g., with high student response rates for each course at each institution) are particularly rare (e.g., Commeford *et al.*, 2021; Sundstrom *et al.*, 2025a). This scarcity is likely due to the challenges of collecting multi-institutional educational data, which include navigating Institutional Review Board (IRB) procedures, standardizing measures across diverse local contexts, recruiting a broad set of participants, and coordinating the logistics of large-scale data collection. At the same time, there is relatively little published guidance on how to design and implement multi-institutional data collection efforts in light of these challenges (e.g., Borrego *et al.*, 2016). Others have detailed guidance on multi-institutional data *analysis* in DBER (e.g., Theobald, 2018; Van Dusen and Nissen, 2019), but guidance on common challenges and potential strategies related to data *collection* across institutions remains underdeveloped in the DBER literature.

This essay aims to offer actionable guidance for DBER researchers seeking to collect multi-institutional data. To do so, we predominantly draw from multi-institutional research in the medical sciences, medical education, clinical research, and psychology, given the limited DBER- and science education-specific literature on these topics. We also describe an example of these strategies applied to a multi-institutional physics education research project where we collected three sources of data (concept inventories, social network surveys, and classroom observations) from 31 introductory physics instructors at 28 institutions in the United States.

## CHALLENGES AND STRATEGIES

In this section, we identify and discuss four major challenges associated with multi-institutional data collection and propose actionable strategies to address them (Table 1). These challenges are not intended to be exhaustive, but rather to highlight several common and consequential issues that arise when doing multi-institutional DBER work. We emphasize that for each challenge, multiple strategies are available, each involving distinct affordances and tradeoffs related to researcher and participant workload, standardization across sites, flexibility for local contexts, and the quality of the resulting data. Rather than suggesting one definitive approach, we encourage researchers to consider which affordances and tradeoffs are most important for their particular study.

***Insert Table 1 approximately here***

## Navigating Institutional Review Board Procedures

Obtaining IRB approval for multi-institutional studies can be complex, and there are several possible approaches. Below, we present four of these approaches, organized from least to most centralized around the research team's IRB. Selecting among these and other approaches will depend on the research team's priorities and local IRB policies, as well as the scale of the project.

The first option is for the research team to prepare and submit separate IRB protocols at every participating institution, with each institution's IRB reviewing the study in accordance with its own policies (**Strategy IRB-1A**). The primary advantage of this approach is that the IRB protocols can be tailored to the specific policies and contexts of each institution, rather than being adapted to fit a single, standardized model. The research team is also likely familiar with preparing IRB protocols for single-site projects, and participants can be confident that the study has been reviewed and approved through their institution's IRB. The main disadvantage of this model is that separate IRB protocols are quite time-consuming to prepare, as each institution has slightly different policy interpretations (Caulfield *et al.*, 2011). For example, a 19-site education research study of medical residents found that an identical survey-based protocol was reviewed as exempt, expedited, or requiring full board review across institutions (Linden *et al.*, 2019). In this same study, approval timelines also varied widely, averaging one to two months but ranging from one to seven months. As such, the timelines of getting approval can vary by institution, which could delay or entirely prevent some of the data collection (e.g., if a relevant course only runs once every few years). More broadly, multi-site ethics review variability can also introduce additional complexity when protocol amendments are required. Because IRBs have discretion in how they interpret and apply regulations to local contexts, the same protocol may require different modifications across institutions, necessitating separate negotiation and site-specific protocol edits. These site-specific requirements can introduce procedural variability when IRBs impose different consent language, study procedures, or other constraints on otherwise identical study designs (Caulfield *et al.*, 2011; Green *et al.*, 2023).

Another approach is to structure the work as collaborative research, designating the instructors whose courses will be involved in the study as local investigators who are responsible for securing IRB approval at their institutions (**Strategy IRB-1B**). Similar to the first approach, this structure allows protocols to be adapted to the specific policies and contexts of each institution, as local investigators can tailor and interpret study procedures within their own IRB's expectations. This approach also distributes the workload of preparing and modifying protocols across the collaboration rather than centralizing it with the research team. This structure may also allow instructor participants to be more involved in the research project as local investigators, which can increase familiarity with the study protocols and investment in study implementation. Indeed, prior multi-site research in health settings shows that local investigators can translate study protocols to better fit local contexts, troubleshoot implementation challenges, and support participant response rates (Friese *et al.*, 2017). However, this option may place an unnecessary burden on instructors, given the time required to prepare an IRB protocol. Similar to the first option, timelines of approval can also vary by institution and may impact data collection (Linden *et al.*, 2019), and site-specific amendments may introduce variability in the study procedures (Caulfield *et al.*, 2011; Green *et al*., 2023).

A third option is to establish IRB authorization (reliance) agreements, where the researchers submit one project-wide protocol at their institution that includes both instructors and students as participants (**Strategy IRB-1C**). In contrast to the approaches described above, this model centralizes ethical review under a single IRB, with participating institutions formally agreeing to rely on the reviewing IRB. This approach can significantly reduce workload by eliminating duplicative IRB review across study sites, while still providing participating institutions with assurance that the study has undergone appropriate ethical review. However, reliance agreements can also introduce additional complexity. In interviews with IRB staff across 20 single-IRB sites, nearly all respondents described reliance agreements as difficult to negotiate, time-intensive, and highly variable in how responsibilities are allocated across relying institutions (Lidz *et al.*, 2018). A hybrid model, in which reliance agreements are used when feasible but the research team otherwise works with a subset of sites to pursue local IRB approval, may balance some of these tradeoffs. Research teams using this hybrid model should consider the extra time needed to prepare and review individual proposals in their data collection plans (i.e., researchers should submit the individual IRB proposals several months before the data collection is intended to begin at that site).

A final approach is to obtain one project-wide IRB approval from the research team's institution that includes instructors and students at multiple sites as participants and does not require an authorization agreement (**Strategy IRB-1D**). This approach represents the model most centralized around the research team's IRB. This option relies on whether relevant institutions are considered to be "engaged in research," defined as having a key role in designing the research, conducting the research, analyzing and interpreting the results, or obtaining informed consent from human subjects (Secretary's Advisory Committee on Human Research Protections [SACHRP], 2022). Under this model, only the reviewing institution is considered "engaged" in human subjects research, while participating sites are designated as "not engaged" (Freise *et al.*, 2017). In practice, this means that although data are collected across multiple institutions, IRB oversight is centralized, and responsibility for protocol management, consent procedures, and data analysis rests with the reviewing institution and the research team. The advantage of this model is that it reduces the workload of both the research team (i.e., only one protocol is needed) and the participants (i.e., no IRB processes at their institution are necessary). Still, some IRBs may be hesitant to grant "not engaged" status, and interpretations of what engagement means can vary widely (SACHRP, 2022). For example, in seeking IRB determinations from 26 actual or potential participating sites involved in neurosurgical and pharmacological randomized clinical trials, 14 institutions accepted a "not engaged" designation, while the other 12 either required local IRB review or declined participation altogether (Wilson *et al.*, 2014). As such, this approach may not be feasible at all researchers' institutions due to local policies, and instructors may be hesitant to participate in the study without full approval from their own institution. Again, a hybrid model where there is one project-wide IRB protocol at the researchers' institution, but the research team works with some participants to prepare a separate IRB protocol at their institutions, may balance these tradeoffs.

These four approaches differ in the level of required investment and how "engagement" in research is defined and applied, with important implications for credit and authorship. The International Committee of Medical Journal Editors (ICMJE) defines authorship based on four criteria: 1) substantial contributions to the research process (e.g., conceptualization, design, data collection, analysis, or interpretation); 2) involvement in drafting or revising manuscripts for publication; 3) approval of the final version; and 4) responsibility for the accuracy and integrity of the work (ICMJE, 2026). However, what constitutes a "substantial" contribution can be ambiguous, and many actions essential to multi-institutional research may not meet authorship criteria despite being critical to the success of the project. Under Strategy **IRB-1D**, instructors and students at participating sites are treated as participants rather than "engaged in research," minimizing local responsibilities while limiting participant involvement in ways that are unlikely to meet authorship criteria. In contrast, under Strategies **IRB-1A**, **IRB-1B**, and **IRB-1C**, instructors may take on roles beyond participation, including IRB submission and local coordination, and may contribute to study implementation in ways that could qualify for authorship. However, extending authorship to participants raises ethical tensions around confidentiality, as public attribution may conflict with the anonymity typically promised to human subjects. In community-based research, contributors' desires for recognition and anonymity are often balanced when making authorship decisions (Castleden et al., 2010). Because of the variability in IRB procedures and participant involvement, we recommend that authorship decisions in multi-institutional studies be established during IRB development and clearly communicated early in the project among all contributors (International Committee of Medical Journal Editors [ICMJE], 2026).

## Recruiting Participants

Existing single-site and multi-institutional DBER studies often leverage data from convenience samples, such as students at the researchers' institution and/or at institutions with whom the researchers have existing connections (Slater and Slater, 2011). However, convenience sampling may significantly limit the representativeness of the data and, consequently, the scope of inference of the study. That is, convenience samples pose threats to the generalizability of study results to other contexts. One psychology study, for example, found significant literacy and numeracy skill differences between convenience samples of undergraduate students and a comparable portion of the general population (Wild *et al.*, 2022). Similarly, a study of third grade reading achievement demonstrated that non-random site selection can produce biased estimates of the impacts of educational interventions on student outcomes (Bell *et al.*, 2016). Therefore, if the goal is to produce findings that generalize across contexts, multi-institutional DBER studies should move beyond convenience samples and aim to collect data from a variety of institution types.

As recommended in a recently published report (Tipton and Olsen, 2022), we recommend for researchers to first decide the population they aim to make inferences about (e.g., undergraduate students in biology programs at Hispanic-Serving Institutions, first-year engineering students in the United States) and design their recruitment strategy around that population (**Strategy R-1**). Once a target population is identified, there are several strategies for recruiting participants from a diverse set of institutions. First, researchers may seek grant advisory board members with professional networks relevant to the project's sampling goals, such that the advisory board members can provide the research team with relevant contacts to email (**Strategy R-2**). Researchers may also post advertisements for study participation on relevant platforms (e.g., national research societies, professional listservs, and social media forums; **Strategy R-3**). Snowball sampling, where the research team asks current study participants to list anyone else they know who may fit the study eligibility criteria, is another effective recruitment strategy (**Strategy R-4**; Merriam and Tisdell, 2016). Finally, it may be useful for the project team to advertise and host virtual information sessions where they provide an overview of the project and expectations of participants (**Strategy R-5**). Generally, we recommend that researchers begin the recruitment process early (e.g., at least six months before data collection begins) to capitalize on as many of these strategies as possible and to coordinate any logistics (e.g., IRB processes) with participants ahead of time. For certain types of institutions, such as community colleges, researchers may need to plan for even longer recruitment timelines (i.e., up to two years) to allow sufficient time to build rapport and navigate IRB collaboration (Sawtelle, 2026). Researchers should also closely monitor recruitment efforts, for example using a spreadsheet that tracks dates of contact and, if applicable, reasons for declining to participate (**Strategy R-6**; Tipton and Olsen, 2022). Such monitoring ensures that the research team follows up with participants within a reasonable time frame and avoids reaching out to contacts who have already declined participation.

As with single-institution DBER studies, it may be especially difficult to recruit participants at traditionally underrepresented institutions (e.g., minority-serving institutions, community colleges, and two-year colleges) and from traditionally underrepresented backgrounds in science (e.g., low-income students; Kanim and Cid, 2020). Instructor participants are also likely to be DBER researchers themselves, which may bias findings (Andrews *et al.*, 2011). We recommend that researchers be intentional about recruiting instructors at a variety of institution types, such as by personally sharing study advertisements with department heads at underrepresented institutions that they can send to instructors in their department (i.e., instead of relying on mass emails that may go unread; **Strategy R-7**).

To support recruitment efforts, we also suggest offering a participation incentive, such as extra credit or monetary compensation, when the resources to do so are available. Incentives have been shown to increase participation rates (Sundstrom *et al.*, 2016; Abdelazeem *et al.*, 2022), which may suffer more in multi-institutional studies than single-site studies because the research team cannot be at each institution to collect the data and send study reminders. Previous research also indicates that cash incentives are more effective than non-monetary or gift card incentives at increasing participation rates (Birnholtz *et al.*, 2002; Kelly *et al.*, 2017). The exact value of cash incentives will be different for every study, but it should be commensurate with the type(s) of data being collected (e.g., time required to participate). Researchers should carefully consider the costs and benefits of study participation when determining a value, as previous work has highlighted that increasing monetary incentives may have diminishing returns on participation rates (i.e., there may be a threshold at which eligible participants view participation as worth the compensation, and any additional compensation does not promote further participation; Kelly *et al.*, 2017). Furthermore, excessive monetary incentives may act as "undue influence," potentially pressuring individuals to participate and raising ethical concerns (Singer and Bossarte, 2006). Undue influence may disproportionately persuade financially vulnerable populations to participate, such as instructors who are more likely to face career instability (e.g., lecturers and adjunct professors) and low-income students, and these populations are often at the traditionally underrepresented institution types we advocate for researchers to recruit. We recommend that research teams consult their grant advisory board and/or local IRB to determine an incentive that boosts participation but does not constitute undue influence to the target population (**Strategy R-8**).

Once the data are collected, it is recommended that researchers revisit the population they initially set out to recruit and consider whether their sample aligns with the intended target population (Tipton & Olsen, 2022). If there is misalignment, researchers may either pursue additional targeted recruitment efforts to better align with the intended population or re-evaluate the scope of the claims they can make based on the sample obtained. For example, if a study aims to make claims about U.S. undergraduate biology students broadly, but includes only students at R1 institutions, researchers should be cautious about extending those claims to students at community colleges or small liberal arts institutions, where similar studies may yield different results.

## Standardizing Measures Across Diverse Instructional Contexts

Another significant challenge is balancing the need for standardized data collection procedures across the study sample (e.g., to allow for data to be aggregated across contexts) with the realities of local variation across institutions (e.g., different academic calendars, course topics, instructors, and physical classroom spaces). To address this challenge, we recommend that researchers standardize data types across sites and centralize data collection infrastructure to the extent possible (i.e., depending on the selected IRB approach; **Strategy S-1**), while also building in participant flexibility (**Strategy S-2**; Louis, 1982). Below, we discuss how this balance may be reached for several common data types in DBER.

**Surveys and Concept Inventories.** We recommend standardizing survey data collection procedures through one of the following options: creating and managing researcher-owned versions of surveys in Qualtrics (or similar online survey software), providing instructors with a standardized template (e.g., spreadsheet) for recording results from their students, or asking instructors to use centralized platforms (e.g., LASSO – Learning About STEM Student Outcomes, n.d.) that allow them to administer assessments and download results in a consistent format. Such processes eliminate the need for instructor participants to score any study surveys or assessments and ensure the same information is collected across contexts. Standardized data formats also significantly facilitate data analysis by reducing the need for extensive data cleaning.

Flexibility can be built into survey administration by allowing instructors to collect survey data within a broader time window (e.g., the first two weeks of class instead of the first day of class) and in a way that least interrupts their course activities (e.g., as an out-of-class assignment instead of an in-class assignment). However, this flexibility means that survey protocols are often implemented under varying conditions. These variations, such as in-class versus out-of-class administration, can affect survey response rates (discussed further in the “Data Collection Logistics” section below) and data quality, as some students complete out-of-class assessments using external resources (Burgess *et al.*, 2026). Moreover, different survey conditions can confound observed outcomes with factors such as the environment in which the survey was taken (e.g., in- versus out-of-class) and survey modality (online versus paper-and-pencil). We recommend that researchers determine a priori the ways that their analysis may be robust to these variations if they offer substantial flexibility in survey administration. For example, when collecting pre- and post-survey data, it may be plausible to assume that student use of external resources is the same on both surveys and so any measured *changes* in student responses may still be valid. Alternatively, researchers may explicitly track survey administration conditions and assess whether responses collected under different conditions can reasonably be combined. For example, researchers can compare response patterns across conditions or account for clustered variation in administration in their modeling approaches.

Regarding specific survey measures, many survey instruments (e.g., social network surveys and attitudinal surveys) apply to students across a wide range of courses and can often be administered without modification. Still, researchers should carefully review all survey items to identify potential mismatches with local instructional contexts that could affect how students interpret or respond to the questions. For instance, if a survey includes an item asking students to report the extent to which outdoor field-based activities support their learning in a biology course, and a participating course does not have a field-based component, this item should be removed for that course or accompanied by a clearly labeled “not applicable” response option.

Collecting multi-institutional concept inventory data is particularly challenging because courses cover a wide array of topics and different concept inventories are used for each topic. In some cases, there are several concept inventories per topic (e.g., in physics, the Force Concept Inventory [Hestenes *et al.*, 1992], Mechanics Baseline Test [Hestenes and Wells, 1992], and Force and Motion Concept Inventory [Ramlo, 2008] all cover introductory mechanics). This variety presents a tradeoff between accurately measuring conceptual learning and maintaining comparability of results across courses. One option is to require that all participating courses use the same concept inventory, which facilitates direct comparisons of data across courses and institutions. However, this approach may reduce measurement validity if students are assessed on content that is not covered in their course. Researchers may restrict study participation to only include courses that cover the topics within the selected concept inventory, but this may substantially limit the pool of eligible study participants, particularly in disciplines where there is not a single standard curriculum. For example, introductory biology courses may emphasize different subdisciplines (e.g., cellular and molecular biology, ecology, genetics), leading to substantial variation in content coverage (Heil *et al.*, 2024).

An alternative approach is to allow instructors to select a research-validated concept inventory that best aligns with their course topics. Although this introduces heterogeneity in the specific instruments used, it improves measurement validity by assessing relevant topics and provides some instructor flexibility. Moreover, challenges related to comparability can be addressed analytically, as quantitative methods such as meta-analysis are well-suited for comparing student learning gains across different concept inventories (e.g., Freeman *et al.*, 2014). When allowing instructors to choose a concept inventory, research teams should either be prepared to provide guidance to the instructors (especially those who are not familiar with DBER) or constrain selection to a set of approved instruments that align with the courses being studied and the research aims. For the former, this process may involve identifying commonly used concept inventories, sharing relevant assessment repositories, and pointing instructors to other support resources (e.g., PhysPort; McKagan *et al.*, 2020). Researchers may disseminate these resources via email, individual meetings with participants, and/or information sessions with participants.

**Classroom observations.** Observations at scale are often conducted live and in person with the aim of characterizing student and instructor behaviors (e.g., Stains *et al.*, 2018; Commeford *et al.*, 2021). There are two approaches to conducting in-person classroom observations at scale. First, the research team may travel to the study sites themselves to conduct the observations (as in Commeford *et al.*, 2021). This approach has the benefit of standardization (e.g., if a structured observation protocol is applied), as there is a consistent set of observers present at all sites included in the study. One tradeoff, however, is that these in-person observations are time- and resource-intensive for the research team. A second approach is to train observers local to each site (e.g., research collaborators and/or instructor participants) and have them conduct the observations (as in Stains *et al.*, 2018). This model alleviates some of the time and resource demands of the first approach; however, the observations may be less standardized because they are conducted by different observers.

Another way to collect multi-institutional classroom observation data is to ask instructors to video record their class sessions and send them to the research team. Similar to the second option above, this approach reduces the time and resources required for the research team to travel to the participating sites. One possibility is to ask instructors to use their own equipment (e.g., Zoom on their computer or an integrated classroom camera) to record class sessions, but this procedure may sacrifice standardization and data quality if there is a wide range in audio and video quality across courses. An alternative approach, therefore, is for instructor participants to record their class sessions using standardized recording equipment provided by the research team (i.e., via mail). This option ensures consistent audio and video quality and reduces the possible burden on instructors to find and use their own recording equipment. For either video recording approach, researchers can build in instructor flexibility by allowing instructors to place the camera in a non-disruptive classroom location rather than a fixed location.

Although video recording enables the collection of multi-institutional classroom observations without the need for travel or local training, it can also introduce challenges. Obtaining consent for multi-institutional video-based data collection can be complex, as legal requirements, federal protections under the Family Educational Rights and Privacy Act (FERPA) of 1974, and institutional recording policies vary across states and institutions (Derry et al., 2010). Some state laws and/or institutional policies prohibit recording without the consent of all participants, whereas others permit recording only if non-consenting individuals are not recorded or identifiable. Likewise, state and institutional policies may not always be aligned; in some cases, institutional policies prohibit recording without the consent of all participants, even when state law permits recording with the consent of only one party. These considerations can affect both the feasibility of conducting a study at a given site and how video data can be used and stored, making it essential for researchers to navigate consent procedures in consultation with participants and their institutional IRBs prior to implementation. It can also be difficult to decide what parts of a course to record to obtain a holistic picture of course activities in multi-institutional video-based research. There is substantial variation in course structures across institutions and disciplines (e.g., studio-style courses that integrate lecture, laboratory, and problem-solving versus large-enrollment courses with separate lecture, laboratory, and recitation sections). Rather than specifying a certain type or number of class sessions to observe, researchers may enhance comparability by observing all class sessions that a student would attend in a single week. Flexibility can be incorporated by observing or allowing instructors to record class sessions during any week of the semester.

**Interviews.** Multi-institutional interview studies afford a rich understanding of phenomena across sites (e.g., Ferrare, 2019; Gutzwa *et al.*, 2024). Similar to in-person classroom observations, however, in-person interviews are resource-intensive to conduct when sites are geographically scattered. Online interviews (e.g., using Zoom) may mitigate these challenges and also increase flexibility and accessibility for participants (e.g., for those with mobility- and/or health-related disabilities; McPadden *et al.,* 2023). However, participants may feel reluctant to turn on their cameras during an online interview (Castelli and Sarvary, 2021) or use obscuring Zoom backgrounds (Oliffe *et al.*, 2021), limiting researchers' ability to observe the cues (e.g., gestures, facial expressions) that often support qualitative analyses. Online interviews can also be more prone to interruptions (e.g., due to lost Internet connection) and unnatural conversation pacing (e.g., due to low audio quality or lag time; Oliffe *et al.*, 2021), which may reduce data quality. All of these facets present a particular challenge for online focus group interviews, where several participants must engage in a conversation with one another and take turns speaking. Overall, online interviews comprise a worthy alternative to in-person interviews for multi-institutional work, but there are challenges that should be considered. Though some of the technological difficulties are beyond the control of the research team, researchers may prepare for online interviews by devising instructions for participants (e.g., asking them to use a static background) and reviewing the interviews as they are collected to evaluate interviewer pacing (Oliffe *et al.*, 2021).

Often, the goal of interviews is to understand the meaning that participants have constructed about their specific experiences (Merriam and Tisdell, 2016). Having a degree of familiarity with the local context is often helpful when conducting these interviews; however, in multi-institutional studies, researchers are rarely able to be deeply immersed in each site. This lack of immersion can make it more difficult to interpret participants' lived experiences in relation to their specific environments. As such, interview protocols should be designed to balance consistency across sites while also having enough flexibility to elicit rich, contextually grounded descriptions of the phenomenon under study. A common approach is to use a shared set of core questions that address broadly relevant constructs while also allowing for follow-up probes that enable participants to elaborate on their local context, in what is known as a semi-structured interview approach (Creswell, 2007; Seidman, 2006). Conducting several individual interviews per site may also help the researchers better understand each context. Finally, we advise researchers to use previously established methods to account for differences within and across sites when conducting multi-site interview analysis to enhance study rigor (see Jenkins *et al.*, 2018; McAlearney *et al.*, 2023 for review). These methods include iterative within-site and between-site analyses to identify both context-specific and broadly applicable themes.

**Data Collection Logistics**

A focal challenge in coordinating data collection across different institutions is the variability in academic calendars and instructor circumstances. Courses may run on a semester or quarterly basis or run infrequently (e.g., every other year) and instructors may face unexpected scheduling constraints such as parental leave. We suggest planning for a longer data collection timeline than would be typical for a single-institution study (e.g., multiple semesters or years; **Strategy L-1**) and for data collection timepoints to be structured relative to instructional milestones (e.g., the first week of class) rather than fixed calendar dates (**Strategy L-2**).

Another key aspect of multi-institutional data collection is maintaining communication with participants. To facilitate this process, we recommend that researchers identify the salient timepoints where they will need to contact instructors (e.g., to share a survey link) and prepare email protocols that can be used for all participants (**Strategy L-3**). These protocols allow researchers to copy and paste the required emails quickly and keep communication consistent across participants. Using a mail merge system can further streamline this process by automatically personalizing messages (e.g., inserting names, institutions, or course details) while still relying on standardized templates. Researchers should also track each major communication they have with participants and send follow-ups when necessary.

Relatedly, we suggest keeping a spreadsheet or other electronic management system for monitoring progress and identifying participants who may need reminders or follow-ups (**Strategy L-4**). Such systems should be updated regularly (i.e., daily or weekly) and structured to allow all members of the research team to quickly and easily access information. A detailed list of information that may be helpful to track and example spreadsheets are provided in *Appendix 1, Supplemental Materials*.

Researchers should also anticipate that response rates for multi-institutional work may be lower than those for single-site work (Borrego *et al.*, 2016); however, intensive tracking and follow-up can significantly improve response rates in multi-institutional studies (Nissen *et al.*, 2018; Yeager *et al.*, 2019). We recommend frequently monitoring response rates when possible (e.g., if the research team owns the surveys in Qualtrics) and sending follow-up emails to instructors with low response rates (**Strategy L-5**). These emails may include suggestions to bolster response rates, such as providing additional reminders to the class, allotting in-class time for students to fill out the survey, or offering participation credit. However, researchers should consider possible tradeoffs of these strategies in light of their project goals before recommending them. For example, allotting in-class time for surveys may boost response rates, but reduce instructor flexibility; meanwhile, administering surveys outside of class improves instructor flexibility by not taking time away from course activities, but may lead to lower response rates and/or student use of external resources (Burgess *et al.*, 2026).

## APPLIED EXAMPLE: CHARACTERIZING ACTIVE LEARNING ENVIRONMENTS IN PHYSICS

All authors on this paper collaborated on a national research project titled "Characterizing Active Learning Environments in Physics." Moving beyond comparisons of active learning to traditional, lecture-based instruction (Wieman, 2014; Dancy *et al.*, 2024), we aimed to compare the implementations and impacts of four distinct active learning methods used to teach introductory physics and astronomy (Sundstrom *et al.*, 2025a). We measured the relative benefits of these methods to student learning using pre- and post-semester concept inventory data. We also collected pre- and post-semester social network surveys and classroom video observations to characterize each method. All data collection was coordinated via email between the research team and instructor participants; that is, the research team did not travel to any of the participating sites. Data were collected from 31 total instructors at 28 institutions over the course of three semesters: fall 2023, spring 2024, and fall 2024.

In early 2022, about 1.5 years before data collection started, we worked with our institution's IRB to choose the most appropriate approach to ethical review given that IRB policies and processes are idiosyncratic (Linden *et al.*, 2019). The scope of our project was quite large, spanning many different institutions; therefore, we decided not to use separate IRB agreements at each site to reduce workload. Additionally, our institution does not allow for expedited studies to use authorization agreements. Thus, we secured project-wide IRB approval from our institution that included both instructors and students at multiple sites as participants (**Strategy IRB-1D**). In the protocol, we included clear procedures for obtaining consent at both the instructor and student levels. We adopted a hybrid model where we also offered to coordinate with individual instructors to prepare a separate IRB protocol at their institution if needed. Only one instructor asked to prepare a separate protocol; however, there was not enough time to prepare it ahead of the intended semester of data collection, and this instructor did not participate in our study. We did not arrange an option for instructors to become co-authors of resulting manuscripts because of the structure of our IRB (i.e., the instructors are participants) and because we wanted to preserve the anonymity of the specific institutions included in our study.

We received IRB approval in spring 2022 and started recruiting participants shortly after. As suggested earlier, we first outlined our target population—instructors of first-semester introductory physics or astronomy instructors at a college or university in the United States using one of four named active learning methods (**Strategy R-1**). Our goal was to collect data from at least eight instructors per active learning method, or 32 total instructors. To support recruitment, we had one grant advisory board member for each of the four active learning methods we analyzed (**Strategy R-2**). Each member was part of the original design of the active learning method and/or was connected to a large network of implementers of the method and so was well-positioned to provide us with an initial list of names of possible participants. In addition to contacting instructors on these lists via email, we frequently posted advertisements for study participation on relevant platforms (e.g., the American Physical Society and Facebook groups for method users) and used snowball sampling, where we asked current study participants to share with us the names of any other instructors they knew who were implementing the method (**Strategies R-3 and R-4**). We incentivized instructor participation by offering $1,000 upon successful data collection (**Strategy R-8**). We determined this cash value through discussions with our grant advisory board, who suggested that this amount would ensure broad faculty participation and high quality data (e.g., high survey response rates). We did not provide

compensation for student participation; however, the instructors may have offered their students incentives to complete the surveys. As recommended earlier, we tracked email contact dates and reasons that instructors declined to participate on a spreadsheet to sufficiently spread out email communications and to avoid unnecessarily reaching out to ineligible participants in future semesters (**Strategy R-6**). We implemented these recruitment strategies until summer 2024 (i.e., we continued recruiting throughout the first two semesters of data collection) and successfully recruited 31 instructors, nearly reaching our goal.

Once instructors agreed to participate in the study, we emailed them to ask for course information (e.g., course title and start/end dates) and other information relevant to collecting each data source (e.g., mailing address to send recording equipment; Figure 1). Below, we describe the ways that we standardized measures and coordinated data collection for each of the three data sources.

***Insert Figure 1 approximately here***

## Concept Inventories

We asked instructors to administer a concept inventory as a pre- and post-assessment, once during the first two weeks of the course and once during the last two weeks of the course (**Strategy L-2**). To ensure that our assessment of students' learning reflected the topics covered in their course, we asked participating instructors to choose a research-validated concept inventory that was most relevant to their course learning goals. We also built in flexibility around how they administered the assessment: instructors could administer the concept inventory online or with paper and pencil, within or outside of class time (e.g., by emailing students a survey link to complete at home voluntarily or as a required homework assignment), and with or without an incentive (e.g., extra credit). We allowed the instructors to send us the data in whatever format was convenient for them (e.g., scans of student responses, electronic spreadsheets of student responses, scored student responses; **Strategy S-2**).

About six months before the first semester of data collection, we prepared to provide guidance to instructors for selecting, downloading, and administering the concept inventory (e.g., using PhysPort; McKagan *et al.*, 2020). About three months before the first semester of data collection, we worked with instructors via email to determine an appropriate concept inventory for their course. At this time, we also prepared three email protocols (**Strategy L-3**): one to remind instructors to administer the pre-semester concept inventory (which we sent a few days before their course began), one to remind instructors to administer the post-semester concept inventory (which we sent three weeks before their course ended), and one to obtain the data (which we sent right after their course ended).

26 of the 31 instructors collected pre- and post-semester concept inventory data with at least 40% response rates and/or with at least 30 students with matched responses. Once we received the concept inventory data, we realized that the inconsistent data format across instructors was not easy to clean and analyze. In some cases, instructors also spent a considerable amount of time scoring the assessments they administered (though we did not ask them to do so). Without a centralized data infrastructure for the concept inventories, we also could not monitor and follow up on response rates. As recommended above, future research studies should consider centralizing concept inventory data collection through researcher-owned tools (e.g., Qualtrics) or other existing tools (e.g., LASSO – Learning About STEM Student Outcomes, n.d.).

**Network Surveys**

We standardized the network surveys by using the same prompt from prior work for all courses (**Strategy S-1**; Commeford *et al.*, 2021). The prompt asks students to report the peers in their course with whom they had a meaningful interaction during the last week and includes a checklist of the names of all students enrolled in each course, from which students can select as many of their peers' names as they want. We did not modify the prompt because peer interactions are applicable across instructional contexts. For each course, we used the same list of students on both the pre- and post-network survey for two reasons. First, keeping an up-to-date roster at any point in the semester is difficult and puts a burden on the participating instructors to communicate roster changes with the research team. Second, analyzing network data with the same set of actors is much easier than those with different sets of actors. Similar to the concept inventories, we built in flexibility by allowing instructors to administer the network surveys within or outside of class time, with or without an incentive, and at some point during the first and last two weeks of class.

We created and managed all network surveys on the research team's Qualtrics account to reduce instructor workload and allow us to track response rates (**Strategy S-1**). Each of the two surveys (i.e., pre- and post-semester) for each course was a separate survey with its own unique link that we shared with instructors, who then shared the survey with their students. About three months before the first semester of data collection, we prepared a survey template in Qualtrics that we could readily duplicate for each course (i.e., we only needed to copy and paste each course roster into the survey). At this time, we also created three email protocols (**Strategy L-3**): one to ask instructors for their current course roster (which we sent right before the first day of their course), one to share the pre-survey link with instructors (which we sent a few days later, after we used the roster to create the survey), and one to share the post-survey link with instructors (which we sent three weeks before the end of the course). After sending each of the pre- and post-survey links, we monitored the incoming responses on Qualtrics and emailed instructors with low response rates to suggest that they remind students to fill out the survey and/or provide an additional incentive (e.g., extra credit; **Strategy L-5**).

19 of the 31 instructors collected pre- and post-semester network surveys with at least 50% response rates. Post-survey response rates were significantly lower than pre-survey response rates, and instructors who administered the network survey in class exhibited much higher response rates than those who administered it outside of class. Future multi-institutional DBER studies, therefore, may need to explicitly prompt instructors to provide additional reminders and/or participation incentives on post-surveys and surveys administered outside of class (Nissen *et al.*, 2018).

**Classroom Observations**

Existing literature suggests that four classroom observations are necessary to accurately characterize instructional style (Stains *et al.*, 2018). To balance this threshold with minimizing instructor burden and standardizing classroom observations across courses, we asked instructors to video record the entirety of three consecutive, typical (e.g., not exam days) class sessions (**Strategy S-1**). We built in flexibility by allowing instructors to record their three class sessions at any point in the semester and with a camera placement that least interrupted the course activities (though they needed to include both the instructor and some students in the camera's field; **Strategy S-2**). Instructors designated an area of the classroom outside of the camera's view where students who did not wish to be recorded could sit.

During the first semester of data collection, we allowed instructors to record using their own recording equipment (e.g., Zoom on their computer) or offered to send them a camera via mail. The videos recorded using instructors' own equipment were not as high quality as those recorded using the mailed cameras (e.g., muffled audio and limited camera field). In the remaining two semesters of data collection, therefore, we required participants to use recording equipment that we mailed to them (**Strategy S-1**).

Three months before data collection began, we purchased 15 cameras (we used the Akaso EK7000 action cameras) with accessories (i.e., batteries, battery chargers, tripods, and memory cards), and packaging materials and postage for 32 instructors (i.e., our target sample size). At this time, we also prepared a short step-by-step guide about how to use the cameras and two email protocols (**Strategy L-3**): one to notify instructors that the camera was shipped to them and provide them with the user guide (which we sent right after we shipped the camera during the first few weeks of the course) and one to remind instructors to collect their recordings and send them back to the research team (which we sent about three weeks before the end of the course). To reduce instructor burden, we included postage and a labeled return package in our initial camera shipment so that instructors only needed to put the cameras into the return package and drop it off at their local post office once they finished recording. All cameras that we mailed to instructors were successfully received by them and returned to us.

30 of the 31 instructors recorded three full class sessions (one of the instructors only recorded one full class session). Upon receiving the observation data, we learned that the varying course structures across institutions coincided with different types of class sessions (e.g., lectures, recitation sections, and laboratory sections) being recorded. We asked each instructor to record three of the same type of class session, but not all instructors chose to record the same type and some instructors recorded class sessions where the active learning method was not prevalent (e.g., they recorded lecture sections when the method was centered around the worksheets implemented during recitation sections). Thus, we recommend for future researchers to collect more consistent observation data across courses, for example by asking instructors to record all class sessions that a student would attend in one week.

Upon successful collection of at least one of the three data sources at the end of their course, we asked instructors to provide us with a W-9 form, which is required by the United States Internal Revenue Service if the sum of one's compensation from a single vendor (in this case, the research team's institution) is at least $600 in one year. We sent the instructors their compensation via physical check or our institution's electronic payment system and concluded their participation in the study.

## CONCLUSION

Based on existing multi-institutional research in the medical sciences, clinical research, and psychology, we have provided strategies for collecting multi-institutional data in DBER studies and described an example of some of these strategies in a multi-institutional physics education research project. The proposed guidance focuses on combatting four common challenges: navigating IRB procedures, standardizing measures across different local contexts, recruiting a broad set of participants, and coordinating the logistics of large-scale data collection. Several strategies are possible for each challenge; therefore, we advise research teams to consider the tradeoffs of each approach in light of their project goals. In general, we recommend that researchers be intentional about recruitment strategies (e.g., identifying a target population and actively reaching out to folks at traditionally underrepresented institutions in DBER), incorporate flexibility into the data collection procedures to the extent possible (e.g., in how and when instructors administer surveys), and stay organized (e.g., preparing email protocols and tracking communications). Given the many challenges of this type of work, we also recommend that multi-institutional research teams publish their data in open repositories (e.g., Sundstrom, n.d.) when possible (e.g., if full participant de-identification is feasible) to facilitate secondary analyses and allow other researchers to build upon the dataset. We hope that this paper mitigates several big-picture concerns about collecting multi-institutional data, facilitating large-scale education research projects. We look forward to seeing the ways in which the DBER community operationalizes the presented strategies in future studies.

## ACKNOWLEDGEMENTS

We thank Ibukunoluwa Bukola, Ian Olivant, and Bruna Schons Ribeiro for meaningful feedback on this manuscript. This material is based upon work supported by the National Science Foundation under Grant Nos. 2111128, 2224786, and 2514310.

**Figure 1. Data collection timeline that we used in our multi-institutional study "Characterizing Active Learning Environments in Physics." Timelines will vary by project size and scope, and the types of institutions involved.**

**Table 1. Summary of challenges and proposed strategies for collecting multi-institutional data in DBER.**

| Challenges | Strategies |
|---|---|
| Navigating IRB procedures | **IRB-1A**: Research team submits a separate IRB protocol at each participating site<br>**IRB-1B**: An instructor at each site submits their own IRB protocol<br>**IRB-1C**: Research team submits one IRB protocol at their institution, and each participating site's IRB signs an authorization agreement to use the researchers' IRB review<br>**IRB-1D**: Research team submits one IRB protocol at their institution that includes instructors and students at multiple sites as participants, with no IRB procedures from participating sites required |
| Recruiting participants | **R-1**: Determine target population of study<br>**R-2**: Select grant advisory board members with relevant networks<br>**R-3**: Post advertisements on several platforms (e.g., national societies and social media)<br>**R-4**: Leverage snowball sampling<br>**R-5**: Host information sessions<br>**R-6**: Track recruitment efforts (e.g., dates of contact)<br>**R-7**: Personally reach out to instructors at traditionally underrepresented institutions and both DBER and non-DBER instructors<br>**R-8**: Provide appropriate participant compensation without undue influence |
| Standardizing measures across diverse local contexts | **S-1**: Centralize data collection infrastructure to the extent possible<br>**S-2**: Incorporate flexibility in instruments used and administration logistics |
| Coordinating data collection logistics | **L-1**: Anticipate an extended data collection timeline<br>**L-2**: Tie data collection to instructional milestones (e.g., the first week of class) instead of specific calendar dates<br>**L-3**: Create email protocols to maintain consistent communication across participants<br>**L-4**: Keep an up-to-date electronic management system to organize data collection information<br>**L-5**: Send follow-up emails to instructors with low survey response rates |

| | 3 to 18 months before data collection starts | A few days before course starts | First two weeks of the course | Three weeks before course ends | Last two weeks of the course | After course ends |
|---|---|---|---|---|---|---|
| **General** | • Obtain IRB approval<br>• Recruit instructor participants<br>• Send email to instructors asking for course and contact information | | | | | • Send email to instructors asking for W-9 form<br>• Provide compensation |
| **Concept inventories** | • Gather resources and help instructors select concept inventories via email<br>• Prepare email protocols | • Send email reminder to instructor to administer pre-concept inventory | | • Send email reminder to instructor to administer post-concept inventory | | • Send email to instructor to send pre- and post-concept inventory data to research team |
| **Network surveys** | • Prepare survey template<br>• Prepare email protocols | • Send email to instructor asking for course roster<br>• Create pre-survey<br>• Send email to instructor with pre-survey link | • Monitor incoming responses on pre-survey; follow up with instructor if low response rate | • Create post-survey<br>• Send email to instructor with post-survey link | • Monitor incoming responses on post-survey; follow up with instructor if low response rate | |
| **Classroom observations** | • Purchase cameras and mailing materials<br>• Prepare camera user guide<br>• Prepare email protocols | | • Mail camera to instructor<br>• Send email with user guide to instructor | • Send email reminder to instructor to collect video recordings and mail camera back to research team | | |